\documentclass[11pt]{article}
\usepackage{latexsym}
\usepackage{amssymb}
\usepackage{epsf}
\usepackage{amsmath}
\usepackage{graphicx}% Include figure files
\renewcommand{\thefootnote}{\fnsymbol{footnote}}
\begin{document}
\pagestyle{empty}

\begin{flushright}
YITP-12-54 \\
TU-914 
\end{flushright}

\vspace{3cm}

\begin{center}

{\bf\Large Supersymmetry-Breaking Nonlinear Sigma Models}
\\

\vspace*{1.5cm}
{\large 
Takumi Imai$^1$, Izawa K.-I.$^{1,2}$, and Yuichiro Nakai$^3$
} \\
\vspace*{0.5cm}

$^1${\it Yukawa Institute for Theoretical Physics, Kyoto University, Kyoto
 606-8502, Japan}\\
\vspace*{0.3cm}
$^2${\it Kavli Institute for the Physics and Mathematics of the
 Universe,\\ University of Tokyo (WPI), Kashiwa 277-8583, Japan}\\
\vspace*{0.3cm}
$^3${\it Department of Physics, Tohoku University, Sendai 980-8578,
 Japan}\\
\vspace*{0.5cm}

\end{center}

\vspace*{1.0cm}

\begin{abstract}
{
\pagestyle{plain}

We consider a novel class of
constraints on chiral superfields to obtain supersymmetric nonlinear
sigma models in four spacetime dimensions,
which strictly combine the internal symmetry breaking
with spontaneous supersymmetry breaking.
The resultant massless modes
can be exclusively Nambu-Goldstone bosons without their complex partners and the
goldstino that is charged under the internal symmetry.
The massive modes show a peculiar relation among their masses and the
scales of symmetry breakings.

}
\end{abstract} 

%%%%%%%%%%%%%%%%%%%%%%%%%%%%%%%%%%%%%%%%%%%%%%%%%%%%%%%%%%%%%%%%%%%%%%%%%%%%
\newpage
\baselineskip=18pt
\setcounter{page}{2}
\pagestyle{plain}
\baselineskip=18pt
\pagestyle{plain}

\renewcommand{\thefootnote}{*\arabic{footnote}}
\setcounter{footnote}{0}

\section{Introduction}

If supersymmetry (SUSY) is realized in nature, it must be broken at some energy scale.
Models of dynamical SUSY breaking have a possibility
to explain smallness of the breaking naturally
\cite{Witten:1981nf}.
These models have some internal (global or gauge) symmetries
at least in the UV region.
Below the scale of strong dynamics,
the SUSY breaking field in an effective theory often appears
as a singlet under all the low-energy symmetries except for $U(1)_R$.
That is, the effective theory can be described by an O'Raifeartaigh-type model
\cite{O'Raifeartaigh:1975pr}.
However, it may be of interest to keep some internal symmetries,
under which the SUSY breaking field is charged, even below the dynamical scale.
In this case, the massless goldstino due to the SUSY breaking is also
charged under these symmetries
\cite{Brignole}.

Effective theories with spontaneously broken global symmetries may be
described by nonlinear sigma models (NLSM) in a general manner.
In the non-SUSY case, one of the ways to obtain a NLSM
is to impose an algebraic constraint on the fields in
a multi-component scalar field theory.
In particular, a NLSM describes Nambu-Goldstone (NG) bosons under a
symmetric constraint. For example, the $O(N)$ NLSM in four spacetime
dimensions is given by a symmetric Lagrangian
\begin{eqnarray}
{\cal L}={1 \over 2}\varphi^i \Box \varphi^i - V(\varphi^{i2}),
\end{eqnarray}
with a constraint
\begin{eqnarray}
\varphi^{i2}-a^2=0,
\end{eqnarray}
where $\varphi^i$ is a real field with a vectorial index $i=1,\cdots,N$
under the $O(N)$, $\varphi^{i2} \equiv \varphi^i \varphi^i$, and $a$ denotes
a positive constant.
Owing to the constraint, the $O(N)$ symmetry is broken down to
$O(N-1)$ with the corresponding $N-1$ NG bosons.
A supersymmetric extension of the above construction is
naively achieved by straightforward use of the corresponding
superfields. Namely, we have $O(N)$-invariant K{\" a}hler potential
$K$ and superpotential $W$ with a superfield constraint
\begin{eqnarray}
X^i X^i-a^2=0,
\label{naiveconst}
\end{eqnarray}
where $X^i$ is a chiral superfield.
The above naive extension keeps supersymmetry unbroken in contrast to
the internal $O(N)$ symmetry, which is broken.

In this letter, we consider
more general form of constraints such as
$[(X^i X^i)^m-a^{2m}]^n=0$,
where $m$ and $n$ are positive integers.
We show that such a constraint can strictly combine the internal symmetry breaking
with spontaneous SUSY breaking
\cite{Bajc}.%
\footnote{
We can utilize the constraint Eq.\eqref{naiveconst} to obtain SUSY breaking models with
separate SUSY-breaking fields.
For instance, a simple model is given by the superpotential
$W =  \mu X^i Y^i,$
where $Y^i$ is a chiral superfield
and $\mu$ is a mass scale.
The field $X^i$ is under the constraint, which gives a part of its lowest component a nonzero
expectation value. On the other hand, the $F$-term of a part of $Y^i$ must develop a nonzero expectation value 
by the equation of motion and thus the SUSY is broken.
We note that the corresponding example of UV dynamical model is given by
a vector-like model of SUSY breaking
\cite{Izawa}, where the $X^i$ is provided by the mesonic degrees of freedom. 
}
Let us focus on
the simplest novel constraint
\begin{eqnarray}
(X^i X^i - a^2)^2 = 0.
\label{const}
\end{eqnarray}
To obtain a SUSY breaking model with this constraint, we only need the
field $X^i$ with the canonical Kahler potential and the superpotential
\begin{eqnarray}
W =  \frac{\mu}{2} X^i X^i,
\label{supo}
\end{eqnarray}
where $\mu$ is a real mass scale.
We will analyze the mass spectrum around the vacuum and show that the resultant massless modes
are exclusively NG bosons without their complex partners%
\footnote{In contrast, unbroken SUSY tends to make complex partner
scalars also massless
\cite{Kugo}.}
and the charged goldstino due to the SUSY breaking.\footnote{
In this paper, the internal symmetry is not gauged.
This provides an O'Raifeartaigh-type SUSY-breaking model
without singlets of the internal symmetry.}

The rest of the paper is organized as follows. In the next section,
we will analyze the constraint to show the SUSY breaking in the model,
which suggests a convenient way of changing variables.
In section $3$, the mass spectrum of the model will be investigated.
In section $4$, we will present a corresponding linear model with additional massive modes.
The final section is devoted to brief discussion.

\section{The model}

In this section, we specify our SUSY-breaking nonlinear sigma model
in detail based on Eq.\eqref{const} and Eq.\eqref{supo}.

\subsection{The constraint}

We consider a chiral superfield%
\footnote{We follow the conventions of Ref.\cite{Wess:1992cp}.}
which transforms as a fundamental representation under a global $O(N)$ symmetry:
\begin{eqnarray}
X^i = x^i(y) + \sqrt{2} \theta \psi^i(y) + \theta^2 F^i(y),
\end{eqnarray}
where $i = 1, \cdots, N$ and 
$y = x + i \theta \sigma \bar{\theta}$ is a four-dimensional coordinate.
Let us impose the superfield constraint Eq.\eqref{const}.
This constraint can be solved as follows.
First, we define a chiral superfield
\begin{eqnarray}
Z &=& z + \sqrt{2} \theta \psi_z + \theta^2 F_z \nonumber \\
&\equiv& X^i X^i - a^2.
\label{zdef}
\end{eqnarray}
Then the superfield constraint Eq.\eqref{const} is nothing but
$
Z^2 = 0
$,
which is equivalent to the relation among the components
\begin{eqnarray}
z = \frac{1}{2F_z} \psi_z \psi_z.
\label{zconst}
\end{eqnarray}
We here assume non-vanishing $F_z$, which will be justified retrospectively.
Note that this takes the same form as the generic constraint for
the goldstino superfield derived in
Ref.\cite{Komargodski:2009rz}.
In terms of the component fields of $X^i$, this relation is rewritten as
\begin{eqnarray}
x^i x^i - a^2 = 2 (2 x^i F^i - \psi^i \psi^i)^{-1} (x^j \psi^j )^2.
\label{comconst}
\end{eqnarray}

\subsection{SUSY breaking}

We next investigate the SUSY breaking in our model.
The superpotential is given by Eq.\eqref{supo}.
In the vacuum, the constraint Eq.\eqref{comconst} leads to
\begin{eqnarray}
\langle x^i \rangle^2 = a^2,
\label{vacuumconst}
\end{eqnarray}
while the $F$-term for the field $X^i$ is given by
\begin{eqnarray}
\langle {F^i} \rangle^\dagger = -\mu \langle x^i \rangle,
\label{fterm}
\end{eqnarray}
where the corrections due to the constraint is vanishing under $\langle \psi^i \rangle=0$.
We see that a part of $\langle F^i \rangle$ must be nonzero in the vacuum
due to the constraint and hence SUSY is spontaneously broken in this
model. The claim $\langle F_z \rangle \neq 0$ is also confirmed.

\subsection{Changing variables}

We now provide a convenient change of variables to be adopted in the following analyses.
Let us assume without loss of generality that only the chiral superfield $X^1$ in $X^i$ has nonzero expectation values
to satisfy the constraint Eq.\eqref{vacuumconst} and the $F$-term equation Eq.\eqref{fterm} as
$
\langle X^1 \rangle = a + \theta^2 F,
$
where $F = -\mu a$. 
The chiral superfield $Z$ defined in Eq.\eqref{zdef} has an expectation value
$
\langle Z \rangle = 2aF \theta^2
$
in the vacuum.
We expand these fields around the vacuum as
$
X^1 \equiv \langle X^1 \rangle + \tilde{X}^1
$ and
$ Z \equiv \langle Z \rangle + \tilde{Z}$.
Then Eq.\eqref{zdef} results in
\begin{eqnarray}
2 \langle X^1 \rangle \tilde{X}^1 = \tilde{Z} - ({X^{\bar{i}}})^2 - ({\tilde{X}^{1}})^2,
\label{vchange}
\end{eqnarray}
where $\bar{i} = 2, \cdots, N$.
By iterative use of this equation, the variable set can be changed from
$X^i$ to $\tilde{Z}$ and $X^{\bar{i}}$
so as to be valid up to arbitrarily higher-order fluctuation terms of the fields.
This serves to analyze the mass spectrum around the SUSY-breaking vacuum in the next section
and also to construct a possible linear model, which we will provide in section 4.

\section{Mass spectrum}

We now investigate the mass spectrum of the model around the vacuum that breaks both the SUSY and the $O(N)$ global symmetry spontaneously.
The variable $\tilde{X}^1$ can be replaced with $\tilde{Z}$ and
$X^{\bar{i}}$ using Eq.\eqref{vchange} repeatedly:
\begin{eqnarray}
\tilde{X}^1 &=& \frac{1}{2} \langle X^1 \rangle^{-1} \left(\tilde{Z} - ({X^{\bar{i}}})^2 - ({\tilde{X}^{1}})^2 \right) \nonumber \\
&=& \frac{1}{2} \langle X^1 \rangle^{-1} \left(\tilde{Z} - ({X^{\bar{i}}})^2 - \frac{1}{4} \langle X^1 \rangle^{-2} \tilde{Z}^2 \right) + \cdots  \nonumber \\
&=& \frac{1}{2} \langle X^1 \rangle^{-1} \left( \left( 1 + \frac{1}{2} \langle X^1 \rangle^{-2} \langle Z \rangle  \right) \tilde{Z} - ({X^{\bar{i}}})^2  \right) + \cdots  \nonumber \\
&=& \frac{1}{2a} \tilde{Z} - \frac{1}{2} \langle X^1 \rangle^{-1} ({X^{\bar{i}}})^2 + \cdots,
\end{eqnarray}
where the ellipses denote the higher-order terms that do not contribute
to masses of the fields.
In the third equality, we utilized the constraint as $\tilde{Z}^2 = - 2 \langle Z \rangle \tilde{Z}$.
It is thus straightforward to replace $X^1$ in our original Lagrangian with $\tilde{Z}$ and $X^{\bar{i}}$:
\begin{eqnarray}
\mathcal{L} &=& \int d^4 \theta \left[ {X^1}^\dagger X^1 + {X^{\bar{i}}}^\dagger X^{\bar{i}} \right] 
+ \left( \int d^2 \theta \, \frac{\mu}{2} X^i X^i  + h.c. \right) \nonumber \\
&=& \int d^4 \theta \left[ \frac{1}{4a^2}\tilde{Z}^\dagger \tilde{Z} + {X^{\bar{i}}}^\dagger X^{\bar{i}} + \left( \frac{\langle X^1 \rangle^\dagger}{2a} \tilde{Z}
-\frac{1}{2} \langle X^1 \rangle^\dagger \langle X^1 \rangle^{-1} ({X^{\bar{i}}})^2 + h.c. \right) \right] \nonumber \\
&&+ \cdots + \left( \int d^2 \theta \, \frac{\mu}{2}  \tilde{Z}   + h.c. \right),
\end{eqnarray}
where the ellipsis denotes the higher-order interaction terms.
Let us further redefine the superfield $\tilde{Z}$ as $\tilde{Z} \rightarrow 2a \tilde{Z}$
to canonically normalize the field.
Then, we obtain the Lagrangian in terms of the component fields as
\begin{eqnarray}
\mathcal{L}
&=& \tilde{z}^\dagger \Box \tilde{z} - i \bar{\psi}_{\tilde{z}} \bar{\sigma}^m \partial_m {\psi}_{\tilde{z}} + F_{\tilde{z}}^\dagger F_{\tilde{z}} +
(x^{\bar{i}})^{\dagger} \Box x^{\bar{i}} - i \bar{\psi}^{\bar{i}} \bar{\sigma}^m \partial_m {\psi}^{\bar{i}} + (F^{\bar{i}})^\dagger F^{\bar{i}} \nonumber \\
&&+ \left( \frac{1}{2} \mu^2 {x^{\bar{i}}}^2 + \mu F^{\bar{i}} x^{\bar{i}} - \frac{1}{2} \mu {\psi}^{\bar{i}} {\psi}^{\bar{i}} + h.c. \right) + \cdots.
\end{eqnarray}
We may use the equations of motion for $F^{\bar{i}}$ and $F_{\tilde{z}}$
with the restricted scalar mode $\tilde{z}$
eliminated by means of the constraint Eq.\eqref{zconst}.
That leads to
\begin{eqnarray}
\mathcal{L}
&=& - i \bar{\psi}_{\tilde{z}} \bar{\sigma}^m \partial_m {\psi}_{\tilde{z}} 
+ (x^{\bar{i}})^{\dagger} \Box x^{\bar{i}} - i \bar{\psi}^{\bar{i}} \bar{\sigma}^m \partial_m {\psi}^{\bar{i}} \nonumber \\
&&- \mu^2 |x^{\bar{i}}|^2 + \left( \frac{1}{2} \mu^2 {x^{\bar{i}}}^2  - \frac{1}{2} \mu {\psi}^{\bar{i}} {\psi}^{\bar{i}} + h.c. \right) + \cdots.
\end{eqnarray}
The fermion ${\psi}_{\tilde{z}}$ is massless and none other than the goldstino in this model.\footnote{
This corresponds to $\langle x^i \rangle \psi^i = \langle x^1 \rangle \psi^1$ in the original variables.}
The other fermions ${\psi}^{\bar{i}}$ have nonzero masses $\mu$.
Diagonalizing the scalar fields $x^{\bar{i}}$ to the mass eigenstates, we find that half of the fields have nonzero masses $\sqrt{2}\mu$,
while the remainings are massless. These are the $N-1$ NG bosons of the spontaneously broken $O(N)$ symmetry.

We can obtain the same result in a component computation.
Let us restrict ourselves to the bosonic components under $\psi^i=0$.
Then the scalar potential is given by
$
V = \left| \mu x^i \right|^2
$
due to the superpotential Eq.\eqref{supo} in spite of the constraint
Eq.\eqref{vacuumconst}, which can be written as
\begin{eqnarray}
(x^1)^2 = a^2 - (x^{\bar{i}})^2.
\end{eqnarray}
Substituting this expression into the scalar potential, we obtain
\begin{eqnarray}
V = \mu^2 \left( |a^2 - {x^{\bar{i}}}^2 |  +  |{x^{\bar{i}}}|^2 \right).
\end{eqnarray}
We now expand the fields $x^{\bar{i}}$ with their real and imaginary parts as
$
{x^{\bar{i}}} = (\xi^{\bar{i}} + i \eta^{\bar{i}} )/\sqrt{2}
$
where $\xi^{\bar{i}}$ and $\eta^{\bar{i}}$ are real fields.
Then the potential is given by
\begin{eqnarray}
V &=& \mu^2 \sqrt{\left(a^2 - \frac{1}{2} {\xi^{\bar{i}}}^2 + \frac{1}{2} {\eta^{\bar{i}}}^2 \right)^2
+ \left( \xi^{\bar{i}}\eta^{\bar{i}} \right)^2 } + \frac{1}{2} \mu^2 ({\xi^{\bar{i}}}^2 + {\eta^{\bar{i}}}^2 ) \nonumber \\
&=& \mu^2 a^2 + \mu^2 {\eta^{\bar{i}}}^2 + \mathcal{O}\left(\xi^4, \xi^2\eta^2, \eta^4 \right).
\end{eqnarray}
We see that there are no quadratic terms of $\xi^{\bar{i}}$, which
correspond to $(N-1)$ NG bosons, as is expected.

\section{A linear model}

We proceed to consider an example of linear models which is with no constraint and effectively realize our constraint
after integrating out certain massive modes.
The Lagrangian is given by
\begin{eqnarray}
\mathcal{L} = \int d^4 \theta \left[ {X^{i}}^\dagger X^{i} + f(Z, Z^\dagger) \right] + \left( \int d^2 \theta \, \frac{\mu}{2} X^i X^i   + h.c. \right).
\end{eqnarray}
Here, the superfield $Z$ is defined by Eq.\eqref{zdef} and all the
fields $X^i$ are independent degrees of freedom.
That is, the superfield constraint discussed so far is not imposed in
this section.
To obtain a meta-stable vacuum at $z=0$, we assume $f(Z, Z^\dagger)$ to take
a higher-order form in $Z$ and $Z^\dagger$ such as
\begin{eqnarray}
f(Z, Z^\dagger) = \left(\frac{b}{M^4}Z^{\dagger} Z^2 + \frac{c}{M^6} Z^{\dagger} Z^3 + h.c. \right) + \frac{d}{M^6}{Z^{\dagger} }^2 Z^2,
\label{higher}
\end{eqnarray}
where $b$, $c$, and $d$ are real constants and $M$ denotes a mass scale.
As in the previous section, we can change the variables from $\tilde{X}^1$ to $\tilde{Z}$ and $X^{\bar{i}}$ using Eq.\eqref{vchange} repeatedly:
\begin{eqnarray}
\tilde{X}^1 
= \frac{1}{2}{\left<X^1\right>}^{-1}  \tilde{Z}-\frac{1}{8}{\left<X^1\right>}^{-3} \tilde{Z}^2+\frac{1}{16}{\left<X^1\right>}^{-5}\tilde{Z}^3+\cdots.
\end{eqnarray}
Then the Lagrangian is rewritten as
\begin{eqnarray}
\mathcal{L}
&=& \int d^4\theta  \biggl[{Z}^{\dagger}Z+\biggl\{8a^3 \left(\frac{b}{M^4}-\frac{1}{16a^4} \right){Z}^{\dagger}{Z}^2 \nonumber\\
&\hspace{5mm}& +16a^4\left(\frac{c}{M^6}+\frac{1}{32a^6}\right){Z}^{\dagger}{Z}^3 + h.c. \biggr\} +16a^4\left(\frac{d}{M^6}+\frac{1}{64a^6}\right){Z}^{\dagger 2}{Z}^2\biggr] \nonumber \\
&\hspace{5mm}& + \cdots + \left( -\int d^2\theta \, F Z +h.c. \right), 
\label{sgoldstino}
\end{eqnarray}
where we have used
$\mu=-{F}/{a}$ and rescaled the field $Z$ as $Z \rightarrow 2a Z$ to canonically normalize the field.
Note that we here adopt $Z$ rather than $\tilde{Z}$ as a variable.

Hence the scalar potential is given by
\begin{eqnarray}
V &=& K^{-1}_{Z,{Z}^{\dagger}} F^2 \nonumber \\
    &=& F^2\biggl[1- \left\{16a^3 \left(\frac{b}{M^4}-\frac{1}{16a^4} \right) {z}+h.c. \right\}- \left\{48a^4 \left(\frac{c}{M^6}+\frac{1}{32a^6} \right){z}^2+h.c.\right\} \nonumber\\
    &\hspace{5mm}&-64a^4 \left(\frac{d}{M^6}+\frac{1}{64a^6} \right){z}^\dagger{z}+256a^6 \left(\frac{b}{M^4}-\frac{1}{16a^4} \right)^2 \left({z}+{{z}^{\dagger}} \right)^2+ \cdots \biggr],
\end{eqnarray}
where $K_{Z,{Z}^{\dagger}}$ is the derivative of the Kahler potential in Eq.\eqref{sgoldstino}
with respect to $Z$ and $Z^\dagger$.
To have a meta-stable vacuum at $z=0$, firstly we have to cancel out the
linear terms in $z$. Thus we are led to set\footnote{
It is not so special to have such a parameters choice.  What we need is just some non-zero VEV of $x^1$, which is actually determined by the higher dimensional operators. 
}
\begin{eqnarray}
b=\frac{M^4}{16a^4},
\end{eqnarray}
to obtain the masses of two real scalars as
\begin{eqnarray}
m_{\pm}^2= F^2\left\{-64a^4 \left(\frac{d}{M^6}+\frac{1}{64a^6} \right)\pm 96a^4 \left(\frac{c}{M^6}+\frac{1}{32a^6} \right) \right\}.
\end{eqnarray}
The vacuum stability around $z=0$ requires
\begin{eqnarray}
2\left(\frac{d}{M^6}+\frac{1}{64a^6}\right)\pm 3\left(\frac{c}{M^6}+\frac{1}{32a^6}\right)<0.
\end{eqnarray}
When this condition is satisfied, we may integrate out the massive scalars
by using the equation of motion for $z$ to reproduce our nonlinear model.
In fact,
Eq.(\ref{sgoldstino}) yields
\begin{eqnarray}
3\left(\frac{c}{M^6}+\frac{1}{32a^6}\right)F^\dagger_z \left(2zF_z-\psi_z \psi_z \right)+2\left(\frac{d}{M^6}+\frac{1}{64a^6} \right)F_z \left(2z^\dagger F^\dagger_z-\bar{\psi}_z \bar{\psi}_z \right)+\cdots=0,
\end{eqnarray}
where the ellipsis denotes the correction terms, which can be neglected
for $m_{\pm} \rightarrow \infty$.
Thus the decoupling limit of $z$ implies
\begin{eqnarray}
2z F_z -\psi_{{z}}\psi_{{z}}=0,
\end{eqnarray}
which coincides with the constraint Eq.\eqref{zconst}.
With this relation, all the higher-order terms Eq.\eqref{higher} vanish and
the model is reduced to the nonlinear model discussed in the previous sections.

%%%%%%%%%%%%%%%%%%%%%
\section{Discussion}
% \label{sec:conclusion}
%%%%%%%%%%%%%%%%%%%%%

We obviously have various possible extensions of the present nonlinear models
to be investigated. Some examples are in order.
The internal symmetry can be gauged, which results in
gauge interactions of the charged goldstino.
Other symmetry groups and symmetry breaking patterns should be considered.
General description based on nonlinear realization might be constructed
by means of supergroups.
Coupling with supergravity%
\footnote{Nonlinear sigma models coupled with broken supergravity
was considered in Ref.\cite{Yanagida}.}
may be intriguing with or without gauging.
It might be possible to construct dynamical SUSY-breaking models
which realize the nonlinear models at low energy.
We even suspect that simultaneous internal and SUSY breaking
may be of some interest also in realistic particle physics models.

\section*{Acknowledgments}

This work was supported by the Grant-in-Aid for the Global COE Program "The Next
Generation of Physics, Spun from Universality and Emergence" 
and World Premier International Research Center Initiative
(WPI Initiative), MEXT, Japan.
YN is supported by JSPS Fellowships for Young Scientists.
YN would like to acknowledge a great debt to his former home institute,
Yukawa Institute for Theoretical Physics, 
where most of this work was done.
\bigskip

%%%%%%%%%%%%%%%%%%%%%%%%%%%%%%%%%%%%%%%%%%%%%%%%%%%%%%%%%%%%%%%

\end{document}